\documentclass[runinauthor,superscriptaddress,twocolumn]{revtex4}
\usepackage{natbib}
\usepackage[dvipdf]{graphicx}
\usepackage{amssymb}
\usepackage{amsmath}
\usepackage{booktabs}
\usepackage{lscape}
\usepackage{txfonts}
\usepackage{color}
\usepackage{ulem}  %unter- (\uline{important}) und durchstreichen (\sout{wrong})
\begin{document}
\title{Theoretical characterization of excitation energy transfer in chlorosome light-harvesting antennae from green sulfur bacteria}
\author{Takatoshi Fujita}
\email{tfujita@fas.harvard.edu}
\affiliation{Department of Chemistry and Chemical Biology, Harvard University, Cambridge, Massachusetts 02138, USA}
\author{Joonsuk Huh}
\affiliation{Department of Chemistry and Chemical Biology, Harvard University, Cambridge, Massachusetts 02138, USA}
\author{Semion K. Saikin}
\affiliation{Department of Chemistry and Chemical Biology, Harvard University, Cambridge, Massachusetts 02138, USA}
\author{Jennifer C. Brookes}
\affiliation{Department of Chemistry and Chemical Biology, Harvard University, Cambridge, Massachusetts 02138, USA}
\affiliation{Department of Physics and Astronomy, University College London, Gower Street, London WC1E 6BT}
\author{Al\'{a}n Aspuru-Guzik}
\email{aspuru@chemistry.harvard.edu}
\affiliation{Department of Chemistry and Chemical Biology, Harvard University, Cambridge, Massachusetts 02138, USA}

\begin{abstract}
We present a theoretical study of excitation dynamics in the chlorosome antenna complex of green photosynthetic bacteria based on a recently proposed model for the molecular assembly. Our model for the excitation energy transfer (EET) throughout the antenna combines a stochastic time propagation of the excitonic wave function with molecular dynamics simulations of the supramolecular structure, and electronic structure calculations of the excited states. We characterized the optical properties of the chlorosome with absorption, circular dichroism and fluorescence polarization anisotropy decay spectra. The simulation results for the excitation dynamics reveal a detailed picture of the EET in the chlorosome. Coherent energy transfer is significant only for the first 50 fs after the initial excitation, and the wavelike motion of the exciton is completely damped at 100 fs. Characteristic time constants of incoherent energy transfer, subsequently, vary from 1 ps to several tens of ps. We assign the time scales of the EET to specific physical processes by comparing our results with the data obtained from time-resolved spectroscopy experiments.
\end{abstract}

\maketitle

%\begin{document}
%\twocolumn
%\setlength{\baselineskip}{24pt}
\section{INTRODUCTION}
%Chlorosome
The Chlorosome is a light-harvesting antenna found in green sulfur and most of green non-sulfur bacteria. Unlike the antenna complexes from other types of phototrophic bacteria, where the light-absorbing pigment bacteriochlorophyll (BChl) molecules are held by protein scaffolding, in the chlorosome BChls are self-aggregated in multiple supramolecular assemblies. These structures, enclosed in a lipid monolayer, form an ovoid shaped body with the characteristic size ranging from tens to several hundreds of nanometers. Electron microscopy (EM)~\cite{Staehelin1978,Hohmann-Marriott2005,Oostergetel2007} and theoretical studies~\cite{Linnanto2008} support the hypothesis that the BChls are organized into tubular elements -- rolls, while sheet-like aggregates~\cite{Hohmann-Marriott2005,Psencik2004} may coexist. The uncertainty in the structural characterization of chlorosomes stems from the large disorder in aggregates composing the chlorosome. The chlorosome is considered to be the largest and most efficient light-harvesting antenna systems found in nature~\cite{Olson1998,Blankenship2004,Overmann2006,Oostergetel2010,Marriott2011,Orf2013}. This observation motivates a strong interest in the design of artificial light-harvesting systems that uses aggregates of pigment or dye molecules~\cite{Prokhorenko2002,Balaban2005,Roger2008,Miyatake2010,Eisele2012}.

In the light-harvesting complex (LHC) of green photosynthetic bacteria the light energy absorbed by the chlorosome is transferred through two consequent units (the baseplate~\cite{Pedersen2010} and Fenna-Matthews-Olson (FMO) complexes~\cite{Olson2004} in green sulfur bacteria and the baseplate and B808-B866 complexes in green non-sulfur bacteria) to reaction centers, where exciton dissociation occurs. Unlike the chlorosome, which contains BChls $c$, or $e$ depending on the species, the other functional units of the LHC  contain BChls $a$ embedded in a protein environment. 

In this study we focus on the energy transfer properties of the chlorosome from $Chlorobaculum$ $tepidum$ ($C.$ $tepidum$) species of green sulfur bacteria (GSB). Recently, Ganapathy \textit{et al}.~\cite{Ganapathy2009}  have proposed a structure of molecular aggregates composing the chlorosome from mutant $C.$ $tepidum$ bacteria, which synthesize BChl~$d$ chlorosomes. The BChls packing in these chlorosomes is more regular as compared to the wild type of bacteria, thus the intermolecular distance can be estimated with a higher accuracy. In order to reveal molecular alignment in BChl aggregates the authors utilized a solid-state nuclear magnetic resonance technique combined with cryo-EM imaging. Linear dichroism spectra of individual chlorosomes~\cite{Furumaki2011} confirm that the chlorosomes from the mutant bacteria are less disordered than those from the wild type. Additionally, the aggregate structure proposed in~\cite{Ganapathy2009} have been supported by two-dimensional polarization fluorescence microscopy experiments~\cite{Tian2011}. We utilize the aggregate model from~\cite{Ganapathy2009} to calculate excitation dynamics and optical response of $C.$ $tepidum$ assuming that the structural dissimilarity between the mutant and the wild type of bacteria is small and its influence on EET within the chlorosome is negligible. In the model of Ganapathy \textit{et al}.~\cite{Ganapathy2009} (Figure 1a) $syn$-$anti$ monomer stacks (Figure 1b) are basic building blocks of the aggregate. The stacks form rings that are organized into helical cylinders. Finally, several concentric cylinders with the interlayer spacing of the order of 2.1~nm form the main body of the chlorosome.

In our recent letter~\cite{Fujita2012}, we have shown that in the chlorosome the memory effects associated with the environment fluctuations -- dynamic disorder -- assist the exciton diffusion for a broad range of static structural imperfections. We characterized the EET by combining all-atom molecular dynamics (MD) simulations, time-dependent density functional theory (TDDFT) excited-state calculations, and open quantum system approaches. In this paper, we revise the excitation structure model in order to obtain more reliable optical and transport properties. To the sake of consistency we provide a detailed discussion of the model. We analyze the energy spectra of the BChl aggregates and compare them with the results obtained from optical spectroscopy measurements~\cite{Psencik1998,Saga2008,Furumaki2012,Martiskainen2012,Dostal2012}. We estimate characteristic time constants of the energy transfer in the chlorosomes and compare them with the experimental data. Finally, we examine the EET in the chlorosome by modeling fluorescence polarization anisotropy decay. Our simulations allow for a unified description of the coherent and incoherent energy transfer and depict the overall time evolution of EET processes in the chlorosome.

%EET dynamics
The EET in chlorosomes have been studied using various time-resolved spectroscopy techniques~\cite{Causgrove1990,Savikhin1994,Savikhin1995,Psencik1998,Prokhorenko2000,Psencik2002,Psencik2003,Martiskainen2009,Martiskainen2012,Dostal2012}.
For later convenience, we briefly review the characteristic time scales measured for chlorosomes from $Chlorobaculum$ $tepidum$ ($C.$ $tepidum$). One-color transient absorption (TA) measurement in the wavelength region of 750-790 nm has provided four major time constants of 200-300 fs, 1.7-1.8 ps, 5.4-5.9 ps, and 30-40 ps~\cite{Psencik1998}. The authors have assigned the first time to relaxation from higher to lower exciton states, the second and third times to the EET within BChl $c$ aggregates in different energy-transfer steps, and the fourth time component to the EET  from BChl $c$ aggregates to the baseplate. Qualitatively similar results have been obtained by two-pulse photon echo and one-color TA experiments~\cite{Prokhorenko2000}. A recent two-color TA study with 685 nm pump and 758 nm probe has also yielded four time measurements of 120 fs, 1.1-1.2 ps, 12-14 ps, and 46-52 ps~\cite{Martiskainen2012}, while the authors have regarded all  time components as the EET within BChl $c$ aggregates.
More recently, two-dimensional (2-D) electronic spectra have been measured to explore exciton dynamics during the first 200 fs after excitation~\cite{Dostal2012}. The authors have shown that the amplitude of the positive peak decays with an effective time constant of 40 fs and have explained changes in the 2-D spectra by proposing exciton diffusion on a sub-100 fs time scale. The experiments above have suggested that there are several EET processes in the chlorosomes from sub-100 fs to several tens of ps, while interpretation of them is still controversial.

\begin{figure}[!t]
\begin{center}
	\includegraphics[width=8cm]{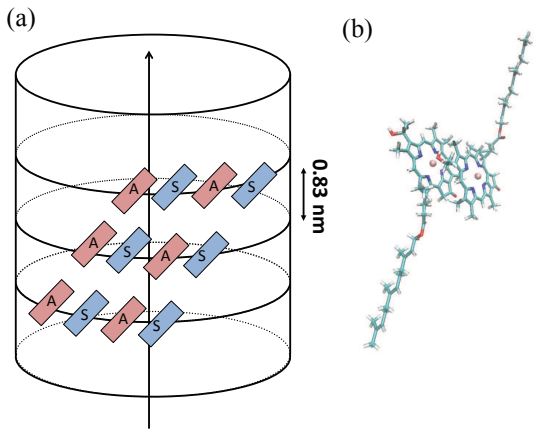}
\end{center}
\caption{The structural model of chlorosomes from the mutant $C.$ $tepidum$ bacteria~\cite{Ganapathy2009}. (a) The schematic model of the molecular assembly. (b) The atomistic structure of the $syn$-$anti$ monomer stacks.}
\end{figure}

% Summary

%Organization
The remainder of this paper is organized as follows:
Section II presents the model of exciton dynamics in the chlorosome.
Section III describes details of the site energy and excitonic coupling calculations.
Section IV A shows spectral densities obtained from MD/TDDFT calculations.
Section IV B presents the optical properties of the roll.
The EET in the chlorosome is discussed in Section IV C to F.
Section IV G summarizes EET time scales in comparison with the experimental data.
In section V we discuss the limitation of the model.
We conclude the study by summarizing our results in Section VI.

\section{THEORY}
% Hamiltonian
We first introduce the excitonic Hamiltonian for the chlorosome.
Each BChl is treated as a two-level system of ground and the $Q_y$ excited state, the excitation energy of which depends on the nuclear configuration.
In the single exciton approximation, the Hamiltonian for the aggregate of BChls can be written as follows:
\begin{align}
H =& \sum_m \epsilon_m(\mathbf{R}) \bigl|m\rangle \langle m\bigr| + \sum_{m\neq n}V_{mn}(\mathbf{R})\bigl|m \rangle \langle n\bigr| \notag \\
     &+ T(\mathbf{P}) + V_{G}(\mathbf{R}),
\end{align}
where $\bigl|m\rangle$ denotes the state where the excitation is localized at $m$-th BChl (site) and all other sites are in the ground states.
$\epsilon_m(\mathbf{R})$ represents an excitation energy of the $m$-th site with the nuclear configuration $\mathbf{R}$, and $V_{mn}(\mathbf{R})$ is a coupling between the electronic transitions of the $m$-th and $n$-th BChls. Here $T(\mathbf{P})$ and  $V_G(\mathbf{R})$ are the kinetic energy and the ground-state potential energy for the nuclear coordinates, respectively. As the next step, we decompose the total Hamiltonian without any assumption of the functional form into a system and bath contributions as:
\begin{equation}
H = H_S + H_{SB} + H_{B},
\end{equation}
\begin{equation}
H_S=\sum_m \left< \epsilon_m \right >\bigl|m \rangle \langle m\bigr|  +
\sum_{m \neq n}  V_{mn}\bigl|m \rangle \langle n\bigr|,
\label{eqn:H_system}
\end{equation}
\begin{equation}
H_{SB}=\sum_m \Delta \epsilon_m(\mathbf{R})\bigl|m\rangle \langle m\bigr| ,
\end{equation}
\begin{equation}
H_{B}=T(\mathbf{P})+V_G(\mathbf{R}),
\end{equation}
where $\Delta \epsilon_m(\mathbf{R})=\epsilon_m(\mathbf{R}) -  \left< \epsilon_m\right >$. The brackets $\left<  \right>$ denote the average over the nuclear configurations, and the Condon approximation was used in order to omit the dependence of $V_{mn}$ on $\bf{R}$.

% dynamical disorder
The time dependence of the site energies $\epsilon_m(\mathbf{R})$ comes from the dependence of their coordinates $\mathbf{R}$ on time $t$.
This dynamic disorder arising from the nuclear motions is determined by the atomistic simulations.
We use MD simulations to generate $\mathbf{R}(t)$ and TDDFT excited-state calculations to obtain $\epsilon (\mathbf{R})$, which is consistent with the models proposed previously~\cite{Damjanovic2002,Olbrich2010,Olbrich2011b,Olbrich2011c,Shim2012}
Thus, in contrast to many studies based on a quantum master equation, this approach can describe the system-bath coupling in complete atomistic detail.
%The magnitude of the dynamical disorder can be quantified by the variance of the site energies  $\left<\Delta \epsilon_m^2 \right>$.

%static disorder
In addition to the dynamic disorder, there is another disorder among excitation energies of the BChls which does not change on the time scales of the exciton dynamics.
The physical origins of this static disorder are due to the existence of different homologues of the BChls~\cite{Huster1990,Borrego1999,Chew2007b}, structural imperfections of supramolecular arrangements, and variations among individual chlorosomes.
Experimentally, the disorder among individual chlorosomes is estimated from low temperature spectral hole burning~\cite{Fetisova1994,Psencik1998} and single molecule fluorescence spectroscopy~\cite{Saga2002,Shibata2006,Furumaki2011,Tian2011}.
It is difficult to predict this static disorder from MD simulations.
Thus we account for the static disorder by introducing random shifts in the average site energy from a Gaussian distribution
$f(\epsilon_m-\left< \epsilon_m \right>)=1/\sqrt{2\pi\sigma^2}\exp{\left( -(\epsilon_m-\left< \epsilon_m \right>)^2/2\sigma^2 \right)}$,
where $\sigma$ is the standard deviation for the static disorder.
The $\sigma$ is assumed to be identical for all BChls.

%\subsection{The time evolution of the exciton dynamics}
%Exciton dynamics
If we treat the nuclear degrees of freedom classically, the wave function of the exciton system is described with the time-dependent Schr$\ddot{\rm{o}}$dinger equation~\cite{May2011}:
\begin{equation}
i\hbar \frac{\partial}{\partial t}\bigl| \psi(t) \rangle= H(t)\bigl| \psi(t) \rangle,
\label{eqn:Schrodinger}
\end{equation}
where $H(t)=H_S+H_{SB}(t)$. In this approach, short MD trajectories are sampled from a full MD trajectory, and the excitonic system is propagated under a unitary evolution for each short MD trajectory. Here in analogy to previously developed models~\cite{Olbrich2011c,Shim2012}, it is assumed that the excitonic system does not affect the bath dynamics. The density matrix of the excitonic system $\rho$ is obtained as an average of these unitary evolutions:
\begin{equation}
\rho(t)=\frac{1}{M}\sum_i^M \bigl| \psi_i(t) \rangle \langle \psi_i(t) \bigr|.
\label{eqn:average}
\end{equation}
A similar approach has been previously applied to small light-harvesting systems such as the FMO complex~\cite{Shim2012,Olbrich2011b,Olbrich2011c} and light-harvesting complexes II (LHII) of purple bacteria~\cite{Damjanovic2002,Olbrich2010}.

%AR(1) Langevin
For a large number of BChls present in the chlorosome, it is computationally unfeasible to calculate site energies of all pigment molecules along an MD trajectory. Here, we take an approximate approach to obtain the site energy fluctuations by solving a set of Langevin equations with white noise:
\begin{equation}
\frac{\partial}{\partial t}\Delta \epsilon_m(t)=-\frac{\Delta \epsilon_m(t)}{\tau_m}+F_m(t),
\label{eqn:Langevin}
\end{equation}
where the stochastic force $F_m(t)$ is determined by $\left <F_m(t) \right>=0$ and $\left < F_m(t)F_m(0)\right>=2\left<\Delta \epsilon_m^2\right>\delta(t)/\tau_m$.
The corresponding energy gap correlation function is from eq~\ref{eqn:Langevin} is
\begin{equation}
\left < \Delta \epsilon_m(t)\Delta \epsilon_m(0)\right>=\left<\Delta \epsilon_m^2 \right>\exp{\left(-\frac{t}{\tau_m} \right)}.
\end{equation}
This procedure gives the same results as the Kubo-Anderson's stochastic model~\cite{Anderson1954,Kubo1954}.
The relaxation time of the bath fluctuations, $\tau_m$, can be determined by integrating autocorrelation functions of the site energies,
$\tau_m=\int_0^{\infty}dt\left <\Delta \epsilon_m(t) \Delta \epsilon_m(0)\right>/\left<\Delta \epsilon_m^2 \right>$.
Thus, the input parameters for the exciton dynamics model $\left< \Delta \epsilon_m\right>$, $\left<\Delta \epsilon_m^2 \right>$, and $\tau_m$, can be obtained from the MD/TDDFT calculations for a single $syn$ and $anti$ BChls dimer considering the symmetry of the roll.

%HSR
In order to check the memory effects, we compare our model with the Haken-Strobl-Reineker (HSR) model~\cite{Haken1972,Haken1973} where the site energy fluctuations do not depend on their memory.
In the HSR model, the correlation functions of the site energies are approximated as
\begin{equation}
\left < \Delta \epsilon_m(t)\Delta \epsilon_m(0)\right >=2\left<\Delta \epsilon_m^2 \right>\tau_m\delta(t).
\label{eqn:Correlator}
\end{equation}
Using eq~\ref{eqn:Correlator}, the time evolution of the density matrix can be obtained as follows~\cite{Haken1972,Haken1973}:
\begin{align}
\frac{\partial}{\partial t}\rho(t)=&-\frac{i}{\hbar}\left[H_S,\rho(t)\right]+\sum_m\gamma_m [ A_m\rho(t)A_m^{\dagger} \notag \\
	                                     &-\frac{1}{2}A_mA_m^{\dagger}\rho(t)-\frac{1}{2}\rho(t)A_mA_m^{\dagger} ],
\end{align}
where $A_m=\bigl|m\rangle \langle m\bigr|$.
Here, a pure dephasing rate for the $m$-th site $\gamma_m$ is given by~\cite{Breuer} $\hbar^2\gamma_m=2\left<\Delta \epsilon_m^2 \right>\tau_m$.

%Hopping model
We also construct a classical hopping model using the following equation~\cite{Ern1972}.
\begin{equation}
\frac{\partial}{\partial t}p_m(t)=\sum_{n}k_{nm}p_n(t)-\sum_{n}k_{mn}p_m(t),
\end{equation}
where the hopping rate from $m$-th to $n$-th sites $k_{mn}$ is given by~\cite{Ern1972},
\begin{equation}
k_{mn}=\frac{V_{mn}^2}{\hbar^2}2\pi \int_{-\infty}^{\infty}d\omega A_m(\omega)F_n(\omega).
\end{equation}
Here, $p_m$ is a population for $m$-th site, and $A_m(\omega)$ and $F_n(\omega)$ are normalized absorption and emission and spectra of the $m$-th and $n$-th sites, respectively.
With the use of eq~\ref{eqn:Correlator}, the absorption and emission lines can be derived as Lorentz functions~\cite{Ern1972} so that the overlap of the absorption and emission spectra becomes:
\begin{align}
2\pi \int_{-\infty}^{\infty}d\omega A_m(\omega)F_n(\omega) \notag \\
=\frac{4(\gamma_m+\gamma_n)}{4(\omega_m-\omega_n)^2+(\gamma_m+\gamma_n)^2},
\label{eqn:Lorentz}
\end{align}
where $\hbar \omega_m=\left< \epsilon_m\right >$.

\section{SITE ENERGY AND EXCITONIC COUPLING CALCULATIONS}
%force fields
We set up a model of three-concentric five-stacked (3x5) rings so that polarizations from several layers can be incorporated into excited-state calculations.
Rings from different layers consist of 60, 80 or 100 BChl molecules, and their radii are 6.1, 8.2, and 10.2 nm, respectively.
The total number of the BChl molecules in the structure is 1200.

%\begin{figure}[!th]
%\begin{center}
%	\includegraphics[width=6cm]{Fig2.eps}
%\end{center}
%\caption{(a) Top view and (b) side view of a model of three-concentric five-stacked rings for site energy and excitonic coupling calculations. A single pair of $syn$ (red) and $anti$ (blue) monomers in the three-concentric five-stacked rings, the site energies and the TrEsp charges of which were obtained from the excited-state calculations, are shown in different colors. \otaka{In (a) and (b), f}\ctaka{F}arnesyl tails of the BChls are not shown for clarity.}
%\end{figure}

%
The force field parameters of the BChl $d$ were prepared as follows:
One $syn$-$anti$ monomer unit was extracted from Ganapathy's original structure, and the Mulliken charges of the dimer were used as partial charges in the force field.
The Mulliken charges were calculated with the Becke three parameter Lee-Yang-Parr hybrid functional~\cite{Becke1993,Lee1988} and 6-31G** basis set.
Other parameters were adopted from the General Amber Force Field~\cite{Wang2004}.

%MD/TDDFT
After the geometry optimization, an MD simulation was performed using a CUDA implementation of NAMD package~\cite{NAMD,Stone2007} with a 1~ns equilibration time and 20~ps production time. The temperature was fixed at 300~K. Then, the $Q_y$ excitation energies were computed for the $syn$ and the $anti$ monomers with the 4~fs time step. The TDDFT calculations were performed using the Q-CHEM quantum chemistry package~\cite{Q-Chem} with the 6-31G basis set and the long-range corrected hybrid functional of Becke~\cite{Chai2008}. Finally, the results were interpolated to 2 fs time steps.

%excitonic couplings
In our previous paper~\cite{Fujita2012}, we employed the point-dipole approximation (PDA) with experimental estimates of the transition dipole moments.
This approximation is strictly applicable only if the intermolecular distance is much larger than the size of the molecules~\cite{Krueger1998,Madjet2006}, which is not the case in the chlorosome where both sizes are comparable.
In order to relax the PDA, we decided to apply transition charges from the electrostatic potentials (TrEsp) method~\cite{Madjet2006} as it offers a compromise between accuracy and computational times.

%TrEsp
In the TrEsp method~\cite{Madjet2006}, the Coulomb interaction between transition densities is approximated as a sum of pairwise interactions of atomic charges.
%\begin{equation}
%\begin{split}
%V_{mn} &=\int d\mathbf{r}_1\int d\mathbf{r}_2 \frac{\rho^{TD}_m(\mathbf{r}_1)\rho^{TD}_n(\mathbf{r}_2)}{|\mathbf{r}_1-\mathbf{r}_2|} \\
%          &=\sum_{I>J} \frac{q^{(m)}_Iq^{(n)}_J}{|\mathbf{R}_I-\mathbf{R}_J|}
%\end{split}
%\end{equation}
%where $\rho_m^{TD}$ are transition densities of monomer $m$, and $q^{m}_I$ are  atomic charges that are determined so that transition densities can be reproduced.
%In the original work CHELPG-BOW methods was used for the fitting procedure~\cite{Madjet2006}.
We employ optimally-weighted charges~\cite{Okiyama2009} to obtain partial atomic charges from transition densities, where Mulliken charges of the transition densities were used as the reference.
A single pair of $syn$-$anti$ stacked monomers in the 3x5 rings were computed quantum-mechanically, and all other molecules were treated as external point charges.
The transition densities were calculated using multi-layer fragment molecular orbital methods~\cite{Mochizuki2005} in conjunction with configuration interaction singles and the 6-31G basis set.
ABINIT-MP(X)~\cite{Mochizuki2005,Nakano2002} program package was used for the excitonic coupling calculations.
Finally, the TrEsp charges of the $syn$ and $anti$ monomers were used to calculate all excitonic couplings in the structure.

\section{RESULTS}
\subsection{Spectral densities}
\begin{figure}[!h]
\begin{center}
	\includegraphics[width=8cm]{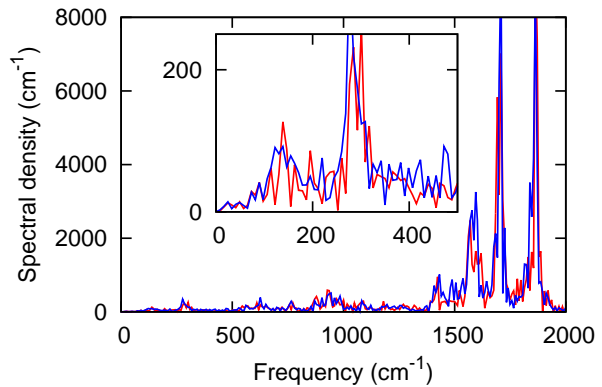}
\end{center}
\caption{Spectral densities for the $syn$ (red) and the $anti$ (blue) monomers with the harmonic prefactor. Low-frequency regions of the spectral densities are shown in the inset.}
\end{figure}

The results of the MD/TDDFT calculations are summarized in Table I.
The calculated average site energies are 15243.0 and 15319.4 cm$^{-1}$ for the $syn$ and the $anti$ monomers, respectively, and the corresponding standard deviations are 532.2 and 556.5 cm$^{-1}$.
Although the $syn$ and the $anti$ monomers do not experience the exact same environment, the difference in the averages and standard deviations of the site energy are small.
The spectral densities for the $syn$ and the $anti$ monomers are shown in Figure 2.
The spectral densities were calculated as a weighted cosine transform of the correlation functions~\cite{Damjanovic2002,Olbrich2011b,Valleau2012}:
\begin{equation}
j_m(\omega)=\frac{\omega}{\pi k_BT}\int_0^{\infty}dt \left<\Delta \epsilon_m(t)\Delta \epsilon_m(0) \right>\cos{(\omega t)}.
\label{eqn:Spectral_density}
\end{equation}
In eq~\ref{eqn:Spectral_density} we used a harmonic prefactor in order to negate the temperature dependence of the classical correlation function.~\cite{Valleau2012} In the obtained temperature-independent spectral density, Figure 2, the strong peaks around 1600 to 2000 cm$^{-1}$ are due to internal vibrational modes of the BChl, while the low-frequency region is ascribed to intermolecular interactions between the BChl molecules.

%Reorganization energy: comparison with bacteriochlorophylls in solvent, implication for fast loss of coherence.
The reorganization energies 658.7 and 719.9 cm$^{-1}$ for $syn$ and $anti$ monomers, respectively, were calculated as $E_{R}=\int^{\infty}_{0}d\omega j(\omega)/\omega $.
These values are larger than those calculated from the spectral densities with the standard prefactor~\cite{Valleau2012} in our previous paper~\cite{Fujita2012}.
Our MD/TDDFT calculations imply that the reorganization energies of the BChl $d$ in the roll is larger than that in solvent (69 cm$^{-1}$)~\cite{Niedzwiedzki2010}.
This large reorganization energy can be a signature of a fast energy relaxation within $Q_y$ exciton states in the roll.

The chlorosome pigments should be distinguished from ideal J-aggregates where the J-band with the largest oscillator strength is also the lowest exciton state (LES)~\cite{Psencik1998,Fetisova1994}. P\v{s}en\v{c}\'{\i}k \textit{et al}.~\cite{Psencik1998} has shown that LESs in chlorosomes from $C$. $tepidum$ are distributed within 760-800 nm range, while the absorption maximum is at 750 nm.
We hypothesise that the relatively large reorganization energy enhances the relaxation from the absorption maximum to the lower exciton states, leading to the energy funneling to the baseplate.

%Comparison with large relaxation time fitted to 2D-experimental spectrum
The relaxation time constants for the $syn$ and the $anti$ monomers were calculated to be 8.80 and 4.51 fs, respectively.
These values are in good agreement with those estimated by the atomistic simulation for the FMO complex~\cite{Olbrich2011b,Shim2012}.
However, the bath correlation time obtained from fitting to the experimental absorption spectrum of the chlorosome~\cite{Dostal2012} is several times longer than the present values.
The reason for this discrepancy can be, for instance, due to the fact that our $ab$ $initio$ model assumes a weak electron-vibrational coupling. Moreover, our model accounts for static disorder in the chlorosome, which was not included in the phenomenological model from Ref.~\cite{Dostal2012}.
We expect that our calculations provide a lower bound estimate for the bath correlations times.

\subsection{Spectra}

To compute the absorption and circular dichroism (CD) spectra we ran simulations of the exciton dynamics using ten-stacked rings (Figure 3a), where each ring is composed of 60 BChls. Initial conditions of the site energy fluctuations $\Delta \epsilon_m(t=0)$ were obtained from the Gaussian distribution with a variance $\left<\Delta \epsilon_m^2 \right>$, and the Euler-Maruyama scheme was applied to solve eq~\ref{eqn:Langevin} in order to obtain $\Delta \epsilon_m(t)$ and define $H(t)$.
Initial conditions of the wave functions were chosen as a localized state on an $anti$ monomer (Figure 3), and wave functions were propagated with the time step of 1.0 fs by diagonalizing $H(t)$ at each step.
The quantum trajectories were averaged over 1000 different realizations of each static disorder.

The absorption and CD spectra were calculated from the exciton propagation by using the following expressions~\cite{Damjanovic2002,Shim2012}:
\begin{align}
I_{Abs}(\omega)\propto &Re\int_0^{\infty}dte^{i\omega t}\sum_{m,n} \left<\mathbf{\mu}_m  \cdot \mathbf{\mu}_n \right> \notag \\
	                 &\times \left\{ \left<U_{mn}(t,0)\right>-\left<U_{nm}^*(t,0)\right>  \right\},
\label{eqn:Absorption}
\end{align}

\begin{align}
&I_{CD}(\omega)\propto \notag \\
		&Re\int_0^{\infty}dte^{i\omega t}\sum_{m,n} \left< ( \mathbf{R}_m-\mathbf{R}_n ) \cdot ( \mathbf{\mu}_m \times \mathbf{\mu}_n) \right> \notag \\
	              &\times \left\{ \left<U_{mn}(t,0)\right>-\left<U_{nm}^*(t,0)\right>  \right\},
\label{eqn:CD}
\end{align}
where $U_{mn}(t,0)$ is an ($m$,$n$) element of the time evolution operator for the $H(t)$.

\begin{figure}[!th]
\begin{center}
	\includegraphics[width=5cm]{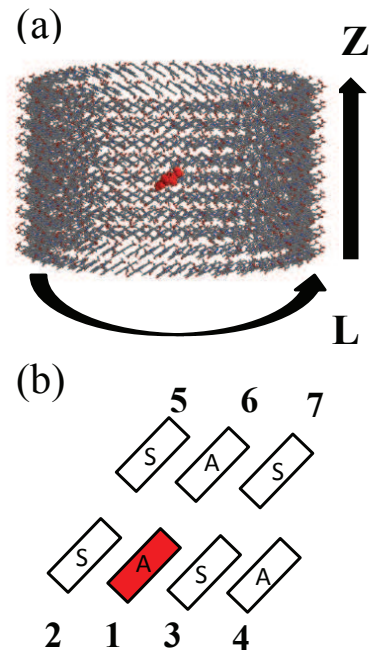}
\end{center}
\caption{(a) A model of ten-stacked rings for the exciton dynamics. (b) The schematic picture of initially excited monomer (red) and its neighboring monomers in the ten-stacked rings. In (a), farnesyl tails of the BChls are not shown for clarity.}
\end{figure}

\begin{figure}[!th]
\begin{center}
	\includegraphics[width=8cm]{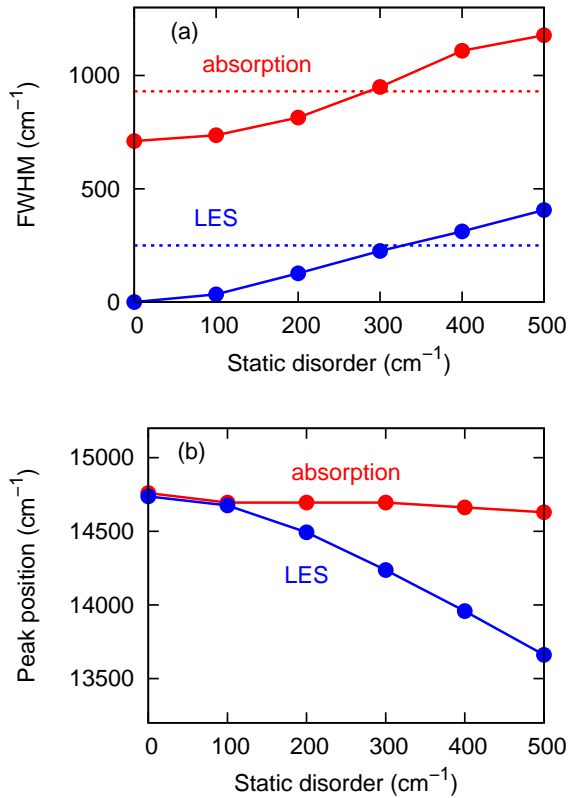}
\end{center}
\caption{(a) Full-widths at half maximum (FWHMs) of absorption spectra (red-solid) and the lowest excited state (LES) (blue-solid) as a function of a standard deviation of the static disorder. Experimental linewidths of the absorption spectra (red-dotted) and the LES (blue-dotted) were taken from Refs.~\cite{Dostal2012} and \cite{Psencik1998}, respectively. (b) Peak positions of the absorption spectra (red) and the LES (blue) as a function of a standard deviation of the static disorder.}
\end{figure}

% Ifentify static disorder
Treating $\sigma$ as a phenomenological parameter, we first estimate its value by comparing the calculated spectra with the experiments. Specifically, we compare our result with spectra measured for the wild type of bacteria because optical properties of the three-point mutant are not well established.
The full width at half maximum (FWHM) of the absorption spectra were calculated as a function of $\sigma$ and compared with the experimental value from Ref.~\cite{Dostal2012}. As both the dynamic and static disorder are included in the exciton dynamics, the linewidth of the absorption spectra computed with eq~\ref{eqn:Absorption} reflects both homogeneous and inhomogeneous broadening.

In addition to the FWHM of the absorption spectra, we further validate our model by comparing the inhomogeneous broadening of the LES.
The inhomogeneous broadening of the LES has been experimentally estimated from the FWHM of the fluorescence spectrum at the temperature of 4 K~\cite{Psencik1998}.
To calculate the corresponding linewidth, the LESs were obtained by diagonalizing the system Hamiltonian with the static disorder, and the FWHM were estimated from a standard deviation of 10000 different realizations of the static disorder.
The resulting FWHM corresponds to the inhomogeneous broadening of the LES.

Figure 4a shows the FWHMs of absorption spectra and the LES as a function of $\sigma$ in comparison with the experiments~\cite{Psencik1998,Dostal2012}. We found that the both FWHMs can be reproduced with $\sigma$ = 300 cm$^{-1}$.

\begin{figure}[!th]
\begin{center}
	\includegraphics[width=8cm]{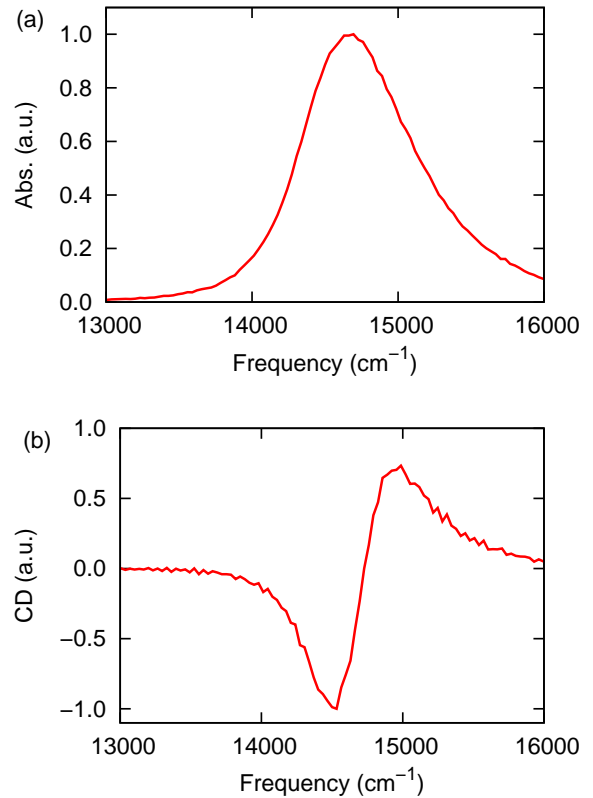}
\end{center}
\caption{(a) Absorption and (b) CD spectra obtained from exciton dynamics with the static disorder $\sigma$ = 300 cm$^{-1}$.}
\end{figure}

%CD
The absorption and the CD spectra with $\sigma$ = 300 cm$^{-1}$ are shown in Figure 5.
The CD spectrum of chlorosomes from the wild type of $C.$ $tepidum$, for example, has two negative and one positive band~\cite{Martiskainen2012}, whereas there is one positive and one negative peak in another experiment~\cite{Saga2008}.
Our CD spectrum agrees with the one reported by Saga \textit{et al}.~\cite{Saga2008} and the recently measured spectra for the three-point mutant~\cite{Furumaki2012}. Our simulated result further supports the supramolecular structure proposed by Ganapathy \textit{et al}.~\cite{Ganapathy2009}

%peak positions
Figure 4b shows the peak positions of the absorption maximum and the LES as a function of $\sigma$.
The peak positions are overestimated compared with the experiments~\cite{Furumaki2012}.
One of the reasons for that is the small basis set in the TDDFT excited-state calculations.
The energy difference between the absorption maximum and the LES is only 23 cm$^{-1}$ without the static disorder.
The absorption maximum is insensitive to $\sigma$, while the LES decreases with increasing $\sigma$.
The energy difference at $\sigma$ = 300 cm$^{-1}$ is 457 cm$^{-1}$ and in reasonable agreement with about 530 cm$^{-1}$ estimated for the wild type chlorosome~\cite{Psencik1998}.

%a role of static disorder?
%In addition to the broadening of the spectra, the static disorder increases the energy difference between the LES and the absorption maximum.
%The LES has a smaller transition dipole moment and thus a longer excited-state life time.
%Experimental study for chlorosomes from $Chlorobium$ $phaeobacteroides$ has shown~\cite{Psencik2003} that most of excitation arrives at the baseplate in 120-130 ps,
% which is remarkably longer than the main decay component of 80-100 ps which were observed at wavelengths within BChl $e$ manifold.
%Authors discussed that their difference is attributed to the contribution of the long-lived LESs, which result in the slowdown of energy transfer to the baseplate.
%The LESs effectively act as a separate spectral pool which provide an energy transfer pathway from the roll to the baseplate, possibly preventing the escape of excitation with a drawback of the slowdown of %the energy transfer.

% relation to experiments
In the following, we discuss how these results are related to experimental observations.
Experimental studies have shown~\cite{Borrego1999,Chew2007b} that absorption spectra of chlorosomes from cells grown at low light intensities exhibit red shift up to 100 cm$^{-1}$.
Borrego \textit{et al}.~\cite{Borrego1999} have found that the red shift is correlated with content of BChl $c$ homologues more methylated at C-8$^2$ and C-12$^1$ positions.
Chew \textit{et al}.~\cite{Chew2007b} have discussed possible effects of those modifications of BChl $c$.
One possibility is that the modifications stabilize formation of larger BChl aggregates which can be assembled into a large chlorosome containing a greater number of BChl $c$.
Another possibility is that the modifications increase disorder in the aggregates and allow a much higher packing density of BChl $c$ in the chlorosomes without causing crystallizations.
Our theoretical calculations show that the peak position of the absorption maximum becomes red-shifted as the static disorder increases. This tendency agrees with the experimental observations. However, the computed shift cannot account for the whole red-shift of the absorption line observed in experiments. Our results are also consistent with the larger linewidths~\cite{Chew2007b} and larger Stokes shifts~\cite{Borrego1999} observed in chlorosomes grown at lower light intensities.
It should be noticed that we treat the static disorder with a simplified model of random shifts in the site energies. In contrast, methylation of BChl molecules will have an influence on both site energies and excitonic couplings by distorting the supramolecular arrangements. It remains to be seen how those modifications induce the disorder in the site energies and excitonic couplings at a molecular level.

%\subsection{Population dynamics}
\subsection{Coherent energy transfer}

%population dynamics and coherence dynamics: selected site, comparison with recent experimental work by Zigmantas and co-workers
Here, we will discuss the coherent energy transfer within the roll.
In order to show how an exciton is delocalized after the initial excitation of a single site, we compute exciton dynamics using eqs~\ref{eqn:Schrodinger}, \ref{eqn:average}, and \ref{eqn:Langevin}.
In addition to the site populations, we present pairwise concurrence between initially excited site and other sites to characterize the role of coherence in the exciton dynamics. The pairwise concurrence~\cite{Sarovar2010} is defined as $2\left| \left<m \left| \rho(t) \right| i \right> \right|$, where $i$ and $m$ refer to the initially excited site and other sites.

\begin{figure*}[!th]
\begin{center}
 	\includegraphics[width=16cm]{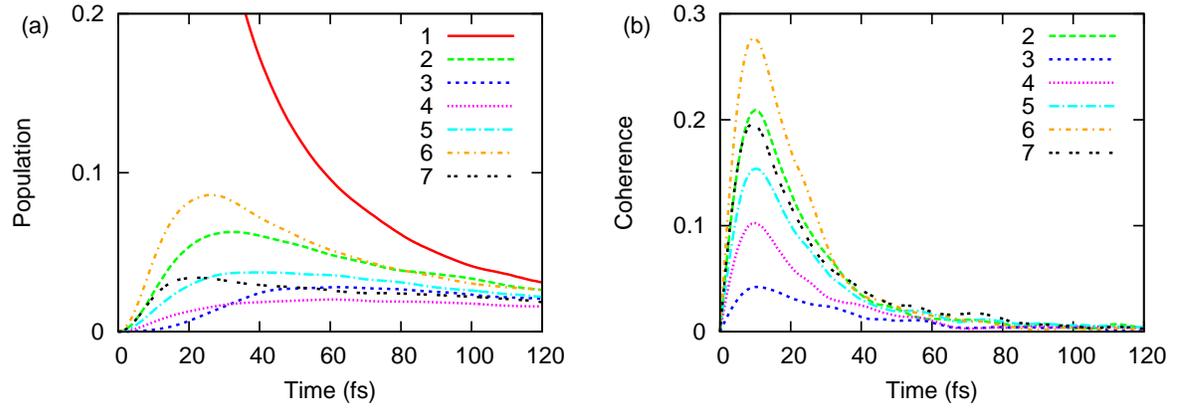}
\end{center}
\caption{(a) Populations and (b) coherences of the selected sites neighboring to the initially excited site, which are obtained with the static disorder $\sigma$ = 300 cm$^{-1}$.
 The site numbers refer to corresponding BChls in Figure 3b.}
\end{figure*}
\begin{figure*}[!bh]
\begin{center}
	\includegraphics[width=16cm]{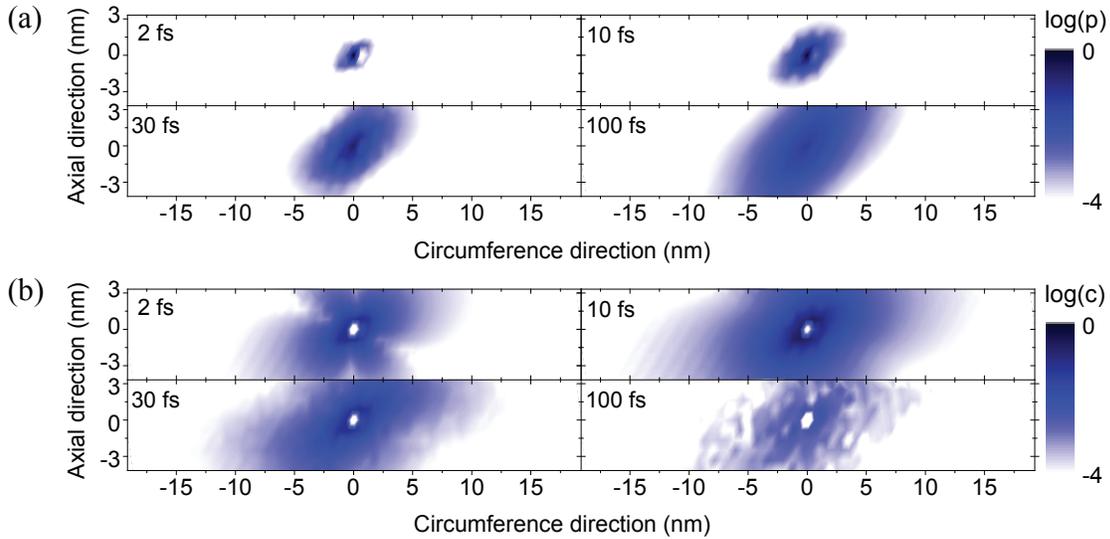}
\end{center}
\caption{Populations (a) and coherences (b) obtained with the static disorder 300 cm$^{-1}$ at selected times after the initial site excitation. In (b), the concurrence of the initially excited site was excluded from the plot as it is same as the population.}
\end{figure*}
\clearpage

% population dynamics of selected site: Timescale. Initial distributions of excitation energies in comparison with experiment.
Figure 6 shows the population and coherent dynamics of selected monomers calculated with $\sigma$ = 300 cm$^{-1}$.
The population of initially excited site decays quickly with a time constant of 25 fs without a oscillatory behavior, which reflects large exciton-vibration couplings.
The coherences increase and decay within 40-50 fs; these values give an estimate of a decoherence time.
The results suggest that the exciton is coherently delocalized within 40-50 fs after the initial excitation of a localized state on a single site, and the coherent propagation is  quickly damped due to the dynamic disorder.

%\begin{figure}[!th]
%\begin{center}
%	\includegraphics[width=8cm]{Fig8.eps}
%\end{center}
%\caption{Coherences obtained with the static disorder $\sigma$ = 300 cm$^{-1}$ at selected times after the initial site excitation. Pink spheres represent Mg atoms of the BChl molecules,
%and the size of the Mg atom reflects the coherent value between the BChl molecule and the initially excited BChl.}
%\end{figure}
In Figure 7, we show the population and coherence dynamics among the roll at selected times.
By comparing Figures 7a and 7b, the coherences spread and are suppressed more rapidly than the populations, while transport directions are the same.
These features are qualitatively similar for all examined values of the static disorder.
The results show that the most of the excitation is distributed over neighboring BChls during the 100 fs after initial excitation, which is comparable to the sub-100 fs exciton diffusion model recently proposed in Ref.~\cite{Dostal2012}.

\subsection{Exciton diffusion within the roll}

%Exciton second moments, isotropic diffusion
To quantify exciton transport on a timescale longer than the decoherence time we compute second moments of the site populations as a function of time.
The second moments of the exciton propagation throughout the circumference (M$_{LL}$) and the coaxial (M$_{ZZ}$) directions (see Figure 3a) calculated with $\sigma$ = 300 cm$^{-1}$ are shown in Figure 8. Both components of the second moment initially scale quadratically and then linearly in time. The transition from the initial ballistic regime to the diffusive regime is observed at 20-30 fs, which is consistent with the time scale of suppression of quantum coherence discussed in the previous section.
In contrast to our earlier study with the PDA couplings~\cite{Fujita2012}, the results show that M$_{LL}$ is comparable to M$_{ZZ}$.
It has been suggested that the $syn$-$anti$ monomer stacks run perpendicular to the symmetry axis in the chlorosome from three-point mutant~\cite{Ganapathy2009}, while they run parallel to the symmetry axis in the chlorosomes from the wild type of bacteria~\cite{Ganapathy2009,Ganapathy2012}.
%Ganapathy et al.~\cite{Ganapathy2012} has recently discussed that the perpendicular and parallel domains can be joined to establish further heterogeneity in supramolecular structure.
Our results may imply that exciton diffusion is insensitive to the supramolecular arrangements, which is relevant to the robust EET.

\begin{figure}[!th]
\begin{center}
 	\includegraphics[width=8cm]{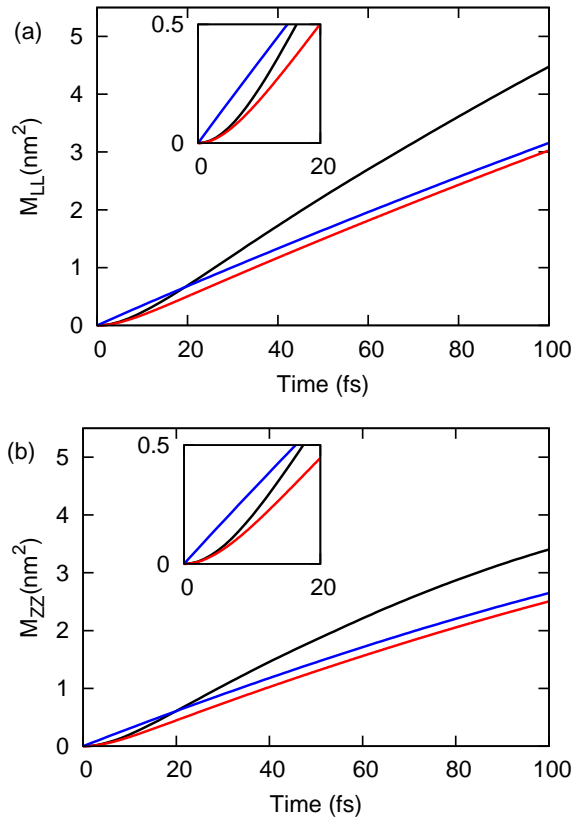}
\end{center}
\caption{Second moments of the exciton populations in the (a) circumference ($M_{LL}$) and (b) coaxial ($M_{LL}$) directions  (see Figure 3a) with the static disorder $\sigma$ = 300 cm$^{-1}$.
Three models compared are Kubo-Anderson (black), Haken-Strobl-Reineker (red), and classical hopping (blue) models.}
\end{figure}

\begin{figure}[!th]
\begin{center}
 	\includegraphics[width=8cm]{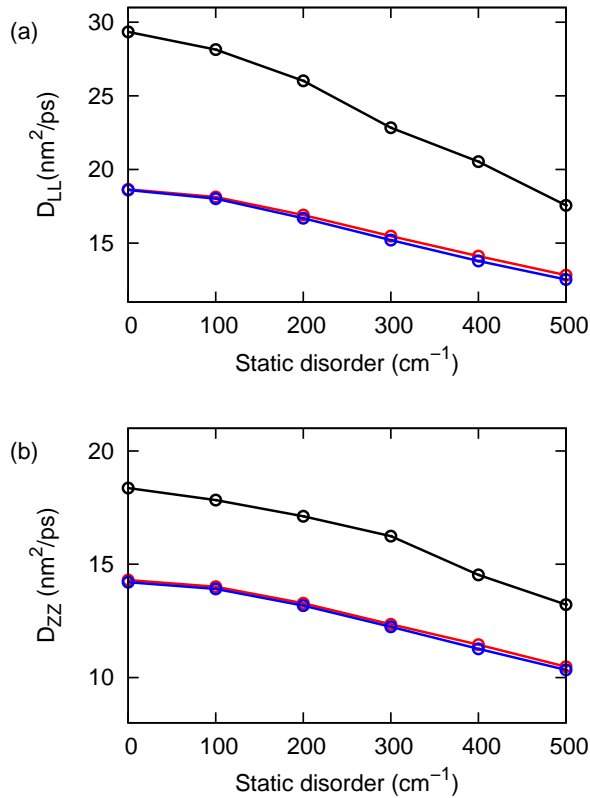}
\end{center}
\caption{Exciton diffusion coefficients in the (a) circumference ($D_{LL}$) and (b) coaxial directions as a function of a standard deviation of the static disorder. Three models compared are Kubo-Anderson (black), Haken-Strobl-Reineker (red), and hopping (blue) models.}
\end{figure}

%comparison
The second moments calculated using the HSR and hopping model are also shown in Figure 8.
The HSR is different from the classical hopping model only within the first 30~fs that correspond to the ballistic exciton propagation. In contrast, the KA model gives a sufficiently faster spread of exciton population over the roll. This effect can be attributed to the non-zero memory of the bath fluctuations as we discussed it in Section II.

%As discussed in Section II, the difference between the KA and the HSR is due to the memory of the bath fluctuations, which is included in the former model.
%This comparison indicates that the memory of the bath fluctuations enhances the diffusion. The HSR and the hopping models have the same slope with the difference only in the initial ballistic propagation up to around 30 fs.
%These results are complementary to our findings that non-Markovianity is also near maximal for the FMO complex~\cite{Rebentrost2009b,Rebentrost2011}.

% The mechanism
The mechanism by which the memory effects enhance the diffusion can be explained by considering site-to-site hopping rates. The hopping rate is determined by the overlap between donor fluorescence and acceptor absorption spectra. If the correlation function of the site energy is a delta function (HSR model), absorption and fluorescence lineshapes become Lorentzian so that the Franck-Condon factor is an overlap of two Lorentz functions (eq~\ref{eqn:Lorentz}). On the other hand, if the correlation function is exponential like the Kubo-Anderson model, lineshapes are closer to the Gaussian~\cite{Ishizaki2010}. It follows that the KA model gives larger overlap between the donor absorption and the acceptor fluorescence spectra. Therefore, the memory effects of the bath fluctuations give larger hopping rates between pigments and thus lead to the larger exciton diffusion.
Note that the memory effects is more pronounced if the bath correlation time is larger than the present value as suggested from the experimental study~\cite{Dostal2012}.

% Diffusion coefficient as a function of the static disorder, time scale of equilibration over single roll.
Finally, we investigate the dependence of the exciton diffusion on the static disorder.
Diffusion coefficients in the circumference and coaxial directions were calculated as a function of $\sigma$ and are shown in Figure 9.
The HSR model gives almost the same diffusion constants as the hopping model: the difference is only in the initial ballistic propagation up to around 30 fs.
The KA models gives about twice as large diffusion coefficients as those from the HSR without the static disorder, while the memory effects become small with increasing $\sigma$.
The time when an exciton arrives at the opposite site of the roll from the initial site excitation can be used as the time scale of exciton equilibration over the single-layer roll, where the exciton populations are equal for all sites.
This time can be calculated as $(R_c\pi)^2/2D_{LL}$, where $R_c$ is the radius of the roll. For the roll of radius $R_c = 6.1$~nm (60 BChl molecules per a ring) with the static disorder of 300 cm$^{-1}$ we obtain the equilibration time to be 8.1~ps.

\subsection{Inter-layer and inter-roll energy transfer}
%\section{Inter-layer and Inter-roll Energy Transfer}
\begin{figure}[!th]
\begin{center}
 	\includegraphics[width=8cm]{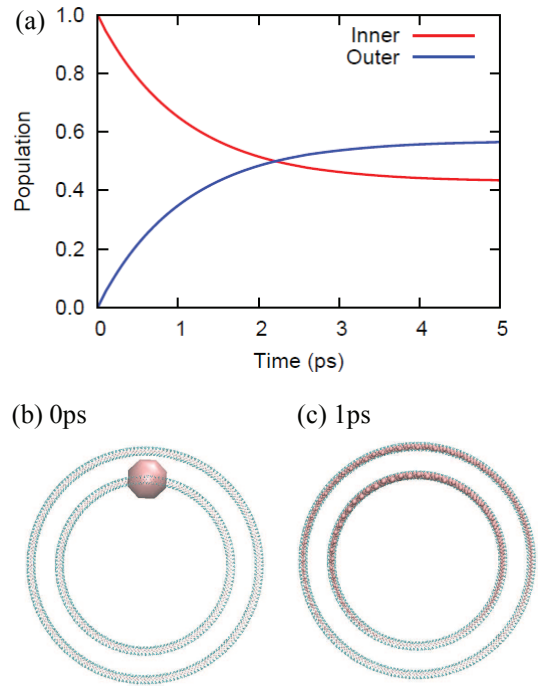}
\end{center}
\caption{(a) Exciton populations of the inner (red) and outer layers (blue) in two-concentric ten-stacked rings. The structure and exciton populations at (a) 0 and (b) 1 ps are also presented.
The pink spheres represent Mg atoms of the BChl molecules, and the size of the Mg atom reflects the population values. In (b) and (c), atoms surrounding the Mg are only shown.}
\end{figure}

So far we have focused on the single-layer roll, yet chlorosomes may consist of several multi-layer rolls as suggested from EM images~\cite{Oostergetel2007,Ganapathy2009}.
Here we explore the EET among different layers and rolls.
We set up two-concentric ten-stacked (2x10) rings and two-roll system consist of 2x10 rings for analysis of the inter-layer and inter-roll EET.
Rings from the inner and outer layers consist of 60 and 80 BChls, respectively.
Their structures and initial conditions for exciton dynamics are shown in Figure 10b and 11b.
In order to speed up our quantum dynamics simulations, we have developed a code for large systems which can propagate wavefunctions in parallel.
The simulation conditions are same as those in the earlier sections, and exciton populations were obtained with $\sigma$ = 300 cm$^{-1}$.

% Inter-layer EET
Figure 10a shows the populations of the inner and outer layers in the 2x10 rings.
%Since the KA model gives results of high-temperature limit and equal populations for all sites in the long time limit, the populations of inner and outer layers are converged to 600/1400 and 800/1400, respectively.
The single exponential fitting of the inner population yields the time constant of 1.1 ps, and double exponential fitting does not improve the fit.
The time of the inter-layer EET is faster than that of the exciton equilibration within the layer.
The time scale of exciton equilibration over the outer layer can be estimated as 14.4 ps by taking into account the diameter of the outer layer.
The result suggests that for a localized initial excitation the exciton is first distributed among different layers with the characteristic time of 1.1 ps and then equilibrated over the roll in 8.1-14.4 ps, respectively for the inner and outer rolls.

\begin{figure}[!th]
\begin{center}
 	\includegraphics[width=8cm]{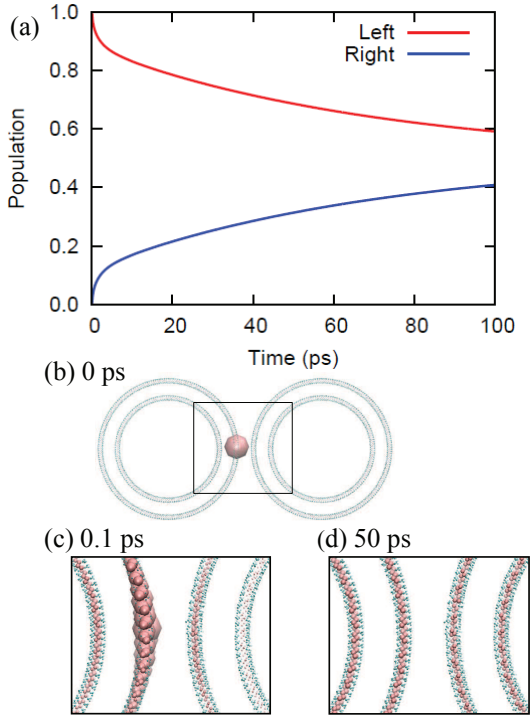}
\end{center}
\caption{(a) Exciton populations of the left (red) and right (blue) rolls in the system composed of two two-concentric ten-stacked rings. (b) Top view of the structure and the populations at 0 ps (the initial condition of the exciton dynamics). The exciton populations at (c) 0.1 and (d) 50 ps are also presented.
The pink spheres represent Mg atoms of the BChl molecules, and the size of the Mg atom reflects the population values. In (b)-(d), atoms surrounding the Mg are only shown.}
\end{figure}

% Inter-roll EET
Figure 11a shows the population dynamics between the two double rolls.
Exponential fitting for the population of the left roll gives two time constants of 1.4 ps (26\%) and 69.8 ps (76\%).
The faster time component of 1.4 ps is compared to that of inter-layer transfer and thus describes the energy transfer between outer layers of the left and right rolls.
Because of the small contact between the rolls, the inter-roll energy transfer becomes slower after the exciton diffuses over the left roll.
The inter-roll EET transfer occurs on major time scale of 69.8 ps, which is much larger than the exciton equilibration over the single roll.
Note that the two-roll system is an extreme structure. In chlorosomes, a roll may be surrounded by other rolls or lamellar layers~\cite{Oostergetel2007}. Thus, the estimated time constants 1.4 and 69.8~ps can be considered as the lower and the upper bounds of the inter-layer energy transfer time scale.

\subsection{Anisotropy decay}
%\section{Anisotropy decay}
%Detail
\begin{figure}[!th]
\begin{center}
 	\includegraphics[width=8cm]{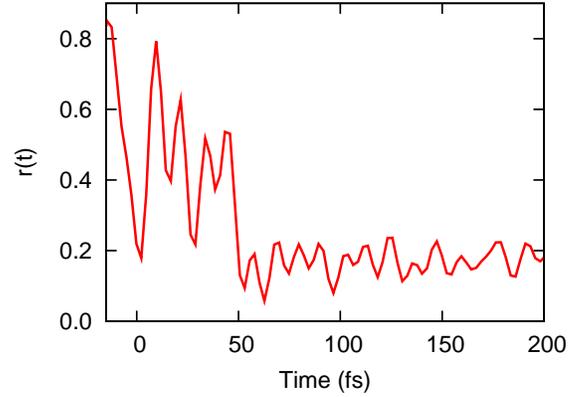}
\end{center}
\caption{The polarization anisotropy decay of the three-staked rings with the static disorder $\sigma$ = 300 cm$^{-1}$.}
\end{figure}

Transient anisotropy decay directly probes the orientational dynamics of transition dipole moments in a system consisting of multiple chromophores and has been used in the analysis of the lifetimes of coherent energy transfer and associated quantum coherence~\cite{Zhu1993,Bradforth1995,Yamazaki2002,Goodson2005,Collini2009,Donehue2011,Parkhill2012}.
Following Ref.~\cite{Parkhill2012}, we calculate the fluorescence anisotropy decay based on the exciton dynamics. In the exciton propagation model we have added the ground state and the off-diagonal system-radiation Hamiltonian, $H_{ij}(t)=\mathbf{E}(t)\cdot \mathbf{\mu}_{ij}$ to eq~\ref{eqn:H_system}.
Here $\mathbf{E}(t)$ is an electric field for an excitation pulse, and $\mathbf{\mu}_{mn}$ is a dipole moment operator.
An ($i$,$j$) component of polarization tensor $\alpha_{ij}$ is defined as the $i$-th component of the dipole moment after applying an electric field polarized in the $j$ direction. The anisotropy decay $r(t)$ can be calculated as $r(t)=3\gamma^2/(45\alpha^2+4\gamma^2)$, where $\alpha$ and $\gamma$ are isotropic and anisotropic tensor invariants of polarization tensor, respectively~\cite{Zare1988}.
An initial condition was chosen as the ground state $\rho= \bigl|G\rangle \langle G\bigr| $, and an 8 fs Gaussian-enveloped electric field of  frequency  14890 cm$^{-1}$ was applied as the excitation pulse. We use the three-stacked rings to calculate the anisotropy decay.

%Ultrafast Depolarization
Figure 12 shows the polarization anisotropy decay for the three-stacked rings structure. The initial value of the anisotropy is close to 0.8, and the residual value is about 0.18.
The anisotropy shows the oscillatory behavior within 50 fs after the excitation pulse, reflecting wavelike motion of the exciton.
Single exponential fitting to the anisotropy profile yields 30 fs as the depolarization time.
The depolarization time can be compared to the decoherence time estimated from the coherences in site basis in Section IV C.
Note that the ultrafast depolarization time of sub-100 fs has not been resolved by earlier time-dependent polarization anisotropy experiments~\cite{Savikhin1994,Savikhin1995,Psencik2003,Martiskainen2012}.
For example, the one-color TA anisotropy in  the wavelength region of 720-790 nm has given a single decay time of 1.7-3.9 ps~\cite{Savikhin1995}.
The lack of the ultrafast component in the earlier experiments is probably due to there limited time resolution.

\subsection{Time scales}
%\section{Time scales}
In this final section, we summarize the various time scales of EET obtained from our simulations. Table II shows the theoretical time scales in comparison with experimental time scales measured for chlorosomes from wild type of $C$. $tepidum$. It should be noticed that the EET time scales may be affected by the presence of the baseplate acting as an exciton drain, which lies outside the scope of this paper.
Exciton dynamics in the whole light-harvesting system, which is composed of the chlorosome, baseplate, and FMO complexes, is being studied in our group, which will be submitted elsewhere.

The experimental studies~\cite{Psencik1998,Prokhorenko2000,Martiskainen2012} have suggested that there are four major time constants in the EET processes.
Here, we compare our results with the time components of 120 fs, 1.1-1.2 ps, 12-14 ps, and 46-52 ps which were obtained with 685 nm pump and 758 nm probe~\cite{Martiskainen2012}, which should have a minimal contribution from the baseplate BChl $a$.
Our simulations show the coherent propagation of excitons for the initial 50 fs after excitation.
The lifetime of this coherent energy transfer can be estimated from the decoherence time and the depolarization time.
The first component of 120 fs can be associated with the coherent EET which describes wavelike motions of the exciton.
The depolarization times of 30 fs is smaller than the first component of the TA experiments.
A possible reason for this is that in terms of the Bloch equation interpretation, the time components of the TA experiments will describe the population decay ($T_1$) time, while the depolarization time includes a pure dephasing time ($T'_2$) as well~\cite{Zhu1993,Yamazaki2002}.
The incoherent EET subsequently takes places after the ultrafast process.
We have obtained 1.1 ps for the inter-layer EET within a single roll and 1.4 ps for the energy transfer between neighboring layers of the double rolls.
The time scale of 1.1-1.4 ps is in good agreement with the second time component of 1.1-1.2 ps; the second component describes the inter-layer energy transfer.
The 8.1-14.1 ps time is comparable to the third component of 12-14 ps, suggesting that the third component reflects the exciton equilibration within a single roll.
The time constant of 69.8 ps corresponds qualitatively to the 46-52 ps.
The forth time component is tentatively assigned as the exciton equilibration over different rolls, while this would not be the case if a roll were surrounded by lamellar layers.

Based on the interpretations above, we speculate on the corresponding time scales of the EET in the chlorosomes based on the structure from the three-point mutant.
The first component describes the lifetime of the coherent energy transfer which largely depends on the exciton-vibration coupling.
Reorganization energies will be similar for the wild type and the three-point mutant as long as the local packing structure is the same.
Therefore, the first component reflecting the coherent energy transfer may be similar for the wild type and the three-point mutant.
We also expect that the second component describing the inter-layer EET is similar for the wild type and the three-point mutant, because the same 2.1 nm spacing was observed for both the wild type and the three-point mutant~\cite{Oostergetel2007,Ganapathy2009}.
The third component describing the energy equilibration over the roll is sensitive to a radius of the roll.
The EM study~\cite{Oostergetel2007} finds that the diameter of the roll of the three-point mutant is larger than that of the wild type.
According to our interpretation and the experimental observation, the third time component of the three-point mutant will be larger than that of the wild type. This could be an experimental verification of our assignment of the experimental time scales.

\section{DISCUSSION}

The model introduced in Section~II utilizes several basic assumptions. Firstly, eqs.~(\ref{eqn:Schrodinger}) and (\ref{eqn:Langevin}) are equivalent to the quantum Langevin equation for the exciton wave function~\cite{Kampen2007} written in a site basis. This equation is strictly applicable in a weak exciton-vibration coupling regime. However, for natural light-harvesting structures an intermediate strength interaction between the electronic excitations and the high-frequency BChl vibrational modes of the range of $1600-2000$~cm$^{-1}$ has been suggested~\cite{Timpmann2004,Ratsep2007,Freiberg2009}. In order to account for this effect polaron models have been proposed~\cite{Damjanovic2002,Meier1997,Kolli2011}. In general, a strong coupling between the exciton and vibrations renormalizes the exciton energy and also reduces its mobility. In Ref.~\cite{Damjanovic2002}, the authors investigated the effects of polaron formation in absorption spectra of LHII of purple bacteria complexes, using a combined molecular dynamics/quantum chemistry approach. They found that the interaction of the excitation with the intramolecular vibrations at about $1670$~cm$^{-1}$ can be described as a weakly coupled polaron. The estimated polaron binding energy was about a half of the exciton bandwidth. In the case of a chlorosome we expect this effect to be weaker. While the polaron binding energy should be about the same, the electronic excitation couplings between the molecules should be stronger due to about 30\% closer packing of BChls. Then, comparing the intermolecular couplings in LHII and the chlorosome within a same model one gets about three times stronger an intermolecular interaction in the latter case. This estimate is consistent with the red shift of the chlorosome and the LHII absorption lines as compared to the corresponding monomers.

Secondly, our classical phonon bath model accounts for dephasing processes only and no feedback of the system on the bath is included. Thus, the long time population dynamics calculated within this model corresponds to an infinite temperature case. However, similar approaches based on classical phonon baths have been compared with reduced density matrix methods~\cite{Ishizaki2011,Berkelbach2012,Aghtar2012}, and they can yield reasonable results compared to exact calculations for a wide range of parameters.

Finally, the supramolecular structure that we use corresponds to mutant bacteria. Thus one can argue that it may be different from the structure of a chlorosome in wild type \textit{C. Tepidum}. For instance in \cite{Ganapathy2012} the authors suggested a different aggregation of BChl $c$ and BChl $d$ in green sulfur bacteria. We expect that the different aggregation can introduce additional corrections to the excitation dynamics that have to be studied in more details.

\section{CONCLUSIONS}
%Summary
In this work, we have characterized the EET in the chlorosome by combining MD simulations, excited-state calculations with TDDFT, and stochastic propagation of an excitonic wave function.
Our study shows that the coherent energy transfer occurs for 50 fs after the initial excitation, and the wavelike motion of the exciton is damped to vanish in 100 fs due to the dynamic disorder.
The incoherent energy transfer subsequently takes place in the characteristic time constants from 1 ps to several tens of ps.
The existence of multiple time scales reflects a hierarchy of the structures in the chlorosomes.
We have interpreted the experimental time scales in terms of the theoretical results and also presented a possible verification of our assignment.

\section*{ACKNOWLEDGMENT}
The authors would like to thank Prof. Huub J. M. de Groot for fruitful discussions and the donation of the $syn$-$anti$ chlorosome configuration template.
We further appreciate St\'{e}phanie Valleau and Prof. Jeongho Kim for very useful discussions. T.F. thanks John Parkhill for advice on the calculations of the anisotropy decay and Yoshio Okiyama for providing a module of the optimally-weighted charge.
T. F., J. H., and A. A.-G. acknowledge support from the Center for Excitonics, an Energy Frontier Research Center funded by the US Department of Energy, Office of Science and Office of Basic Energy Sciences under award DE-SC0001088.
J. C. B. acknowledges support from Welcome Trust UK.
S. K. S. and A. A.-G. also acknowledge Defense Threat Reduction Agency grant HDTRA1-10-1-0046.
Further, A. A.-G. is grateful for the support from Defense Advanced Research Projects Agency grant N66001-10-1-4063, Camille and Henry Dreyfus Foundation, and Alfred P. Sloan Foundation
%\bibliographystyle{achemso}
%\bibliography{roll2}

\clearpage
\begin{table}[!h]
\caption{Results from MD/TDDFT calculations. Averages and standard deviations of the site energies, relaxation times of the site energies, the pure dephasing rates, and reorganization energies are provided for $syn$ and $anti$ monomers. %The relaxation time are shown in fs, while other parameters are given in cm$^{-1}$
}	
\begin{center}
\begin{tabular}{cccccc} \hline
      	      & $\left< \epsilon \right>$(cm$^{-1}$)  & $\sqrt{\left< \epsilon^2 \right>}$(cm$^{-1}$) & $\tau$(fs)& $\gamma$(cm$^{-1}$) & $E_R$(cm$^{-1}$)  \\ \hline
~~$syn$~~&  ~~~15243.0~~~ &  ~~~532.2~~~ & ~~~8.88~~ & ~~~938.8~~ & ~~658.7~(266.5)$^a$~~ \\
~~$anti$~~&  ~~~15319.4~~~ &  ~~~556.5~~~ & ~~~4.51~~ & ~~~525.8~~ & ~~719.9~(298.5)$^a$~~ \\ \hline
\end{tabular}  \\
$^a$Reorganization energies calculated from the spectral densities with the standard prefactor.~\cite{Fujita2012}
\end{center}
\end{table}

\begin{table}
\caption{Energy transfer time scales obtained from our simulations in comparison with the experimental time scales.}
\begin{tabular}{lc} \\ \hline
Theoretical time scales &     \\ \hline
Decoherence time       &  40-50 fs \\
Equilibration over single roll          &  8.1-14.1 ps \\
Inter-layer transfer                    &  1.1-1.4 ps \\
Inter-roll transfer                       &  69.8 ps \\
Fluorescence anisotropy decay            &  30 fs \\ \hline
Experimental time scales &               \\ \hline
One-color TA anisotropy decay (720-790 nm)~\cite{Savikhin1995}    & 1.7-3.9 ps \\
One-color TA (720-790 nm)~\cite{Psencik1998} &  200-300 fs, 1.7-1.8 ps, 5.4-5.9 ps, 30-40 ps \\
Two-pulse photon echo (740-780 nm)~\cite{Prokhorenko2000}  & 140 fs, 1 ps, 10 ps, 280 ps    \\
Two-color TA (685 nm pump, 758 nm probe)~\cite{Martiskainen2012}& 120 fs, 1.1-1.2 ps, 12-14 ps, 46-52 ps \\
2-D spectral line-shape dynamics~\cite{Dostal2012}   &  40 fs$^a$ \\ \hline
\end{tabular} \\
$^a$ Decay time of the amplitude at the position of initial maximum~\cite{Dostal2012}.
\end{table}

\end{document}